\documentclass[11pt,a4paper]{article}

\usepackage[dvips]{graphicx}
\usepackage{calrsfs}
\usepackage{amsmath}

\newcommand{\CL}{\ensuremath{\mathcal{L}}}
\newcommand{\CT}{\ensuremath{\mathcal{T}}}
\newcommand{\CR}{\ensuremath{\mathcal{R}}}
\newcommand{\CA}{\ensuremath{\mathcal{A}}}

\begin{document}
\date{}
\title{Ground state correlations and structure of
odd spherical nuclei}
\author{S. Mishev \thanks{%
Electronic mail:mishev@theor.jinr.ru}, V. V. Voronov \thanks{%
Electronic mail:voronov@theor.jinr.ru} \\
Bogoliubov Laboratory for Theoretical Physics\\
Joint Institute for Nuclear Research \\
6 Joliot-Curie str. \\
Dubna 141980, Russia}

\maketitle

\begin{abstract}
\ \  It is well known that the Pauli principle plays a
substantial role at low energies because the phonon operators are
not ideal boson operators. Calculating the exact commutators between
the quasiparticle and phonon operators one can take into account the
Pauli principle corrections. Besides the ground state correlations
due to the quasiparticle interaction in the ground state influence
the single particle fragmentation as well. In this paper, we
generalize the basic QPM equations to account for both
 mentioned effects. As an illustration of our approach,
calculations on the structure of the low-lying states in $^{131}$Ba
have been performed.
\end{abstract}

\section{Introduction}

In the forthcoming period there will be an increasing activity in
the domain of unstable nuclei studies due to the start of
operation at several major facilities. A theoretical investigation
of odd nuclei far from stability demands to include the ground
state correlation effects.

The quasiparticle-phonon nuclear model (QPM) \cite{Sol92} is
widely used for the description of the energies and fragmentation
of nuclear excitations. The different versions of the QPM
equations for odd spherical nuclei are given in
\cite{Khuong81,Vdo85,GSV88}. It has been shown
\cite{Khuong81,Aliko81} that the Pauli principle plays a
substantial role at low energies, but the ground state correlation
effects were not taken into account in these calculations. From
the other side, the ground state correlations influence the single
particle fragmentation \cite{Sluys83} as they shift the strength
to higher excitation energies.

In this paper, we generalize the basic QPM equations to account
for both mentioned effects. We treat long-range ground state
correlations by including backward-going quasiparticle-phonon
vertices using the equation of motion method \cite{Rowe70} with
explicitly accounting for the Pauli principle. As an illustration
of our approach, calculations of the structure of the low-lying
states in $^{131}$Ba have been performed.

\section{Basic formulae}

We employ the QPM-Hamiltonian including an average nuclear field
described by the Woods-Saxon potential, pairing interactions,
isoscalar particle-hole residual forces in separable form with the
Bohr-Mottelson radial dependence \cite{Bohr75}:

\begin{eqnarray}\label{eq:Hamiltonian}
H =\sum\limits_\tau ^{(n,p)}\{\sum\limits_{jm}(E_j-\lambda _\tau
)a_{jm}^{\dagger }a_{jm}-\frac 14G_\tau ^{(0)}:(P_0^{\dagger
}P_0)^\tau :-\frac 12\sum\limits_{\lambda \mu }\kappa ^{(\lambda
)}:(M_{\lambda \mu }^{\dagger }M_{\lambda \mu }):\}.
\end{eqnarray}

The single-particle states are specified by the quantum numbers
$(jm)$; $E_j$ are the single-particle energies; $\lambda_{\tau}$ is
the chemical potential; $G^{(0)}_{\tau}$  and $\kappa^{(\lambda)}$
are the strengths in the p-p and in the p-h channel, respectively.
The sum goes over protons(p) and neutrons(n) independently and the
notation $\tau = \{n,p\}$ is used.The pair creation and the
multipole operators entering the normal products in
\eqref{eq:Hamiltonian} are defined as follows:
\begin{equation}
P_0^{+}\,=\,\sum_{jm}(-1)^{j-m}a_{jm}^{+}a_{j-m}^{+}, \nonumber
\end{equation}

\begin{equation}
M_{\lambda \mu }^{+}\,=\,\frac 1{\sqrt{2\lambda
+1}}\sum_{jj^{^{\prime }}mm^{^{\prime }}}f_{jj^{^{\prime
}}}^{(\lambda )}\langle jmj^{^{\prime }}m^{^{\prime }}\mid \lambda
\mu \rangle a_{jm}^{+}a_{j^{^{\prime }}m^{^{\prime }}}, \nonumber
\end{equation}
where $f^{(\lambda)}_{jj'}$ are the single particle radial matrix
elements of the residual forces.

In what follows we work in quasiparticle representation, defined by
the canonical Bogoliubov transformation:

\begin{equation}
a^{+}_{jm}=u_j\alpha^{+}_{jm}+\left( -1 \right) ^{j-m}
v_j\alpha_{j-m}.\nonumber
\end{equation}

The Hamiltonian can be represented in terms of bifermion
quasiparticle operators (and their conjugate ones):
\begin{equation}
B(jj^{^{\prime }};\lambda \mu )\,=\,\sum_{mm^{^{\prime
}}}(-1)^{j^{^{\prime }}+m{^{\prime }}}\langle jmj^{^{\prime
}}m^{^{\prime }}\mid \lambda \mu \rangle \alpha _{jm}^{+}\alpha
_{j^{^{\prime }}-m^{^{\prime }}}, \nonumber
\end{equation}

\begin{equation}
A^{+}(jj^{^{\prime }};\lambda \mu )\,=\,\sum_{mm^{^{\prime
}}}\langle jmj^{^{\prime }}m^{^{\prime }}\mid \lambda \mu \rangle
\alpha _{jm}^{+}\alpha _{j^{^{\prime }}m^{^{\prime }}}^{+}.\nonumber
\end{equation}

The phonon creation operators are defined in the two-quasiparticle
space in a standard fashion:
\begin{equation}
Q_{\lambda \mu i}^{+}\,=\,\frac 12\sum_{jj^{^{\prime }}}\{\psi
_{jj^{^{\prime }}}^{\lambda i}\,A^{+}(jj^{^{\prime }};\lambda \mu
)-(-1)^{\lambda -\mu }\varphi _{jj^{^{\prime }}}^{\lambda
i}\,A(jj^{^{\prime }};\lambda -\mu )\},\nonumber
\end{equation}
where the index $\lambda = 0, 1, 2, 3, ...$ denotes multipolarity
and $\mu$ is its $z$-projection in the laboratory system.The
normalization of the one-phonon states reads:

\begin{equation}
\langle|[Q_{\lambda \mu i},Q^{+}_{\lambda ' \mu ' i
'}]|\rangle=\delta_{\lambda \lambda '}\delta_{\mu \mu '} \delta_{i
i'}.\nonumber
\end{equation}

In terms of quasiparticles and phonons the Hamiltonian is rewritten

\begin{equation}
H\,=\,h_0\,+\,h_{pp}\,+\,h_{QQ}\,+\,h_{QB},\nonumber
\end{equation}

\begin{equation}
h_0\,+\,h_{pp}\,=\,\sum_{jm}\,\varepsilon _j\,\alpha
_{jm}^{+}\,\alpha _{jm},\nonumber
\end{equation}

\begin{equation}
h_{QQ}=-\frac 18\,\sum_{\lambda \mu ii^{^{\prime }}}\,{\CA} \left(
{\lambda i i'} \right)\,(Q_{\lambda \mu i}^{+}\,+\,(-)^{\lambda -\mu
}\,Q_{\lambda -\mu i})\,(Q_{\lambda -\mu i^{^{\prime
}}}^{+}\,+\,(-)^{\lambda +\mu }\,Q_{\lambda \mu i^{^{\prime
}}}),\nonumber
\end{equation}

\begin{equation}
h_{QB}\,=\,-\frac 1{2\sqrt{2}}\,\sum_{\lambda \mu ijj^{^{\prime
}}}\, \frac{\pi_{j}}{\pi_{\lambda}} \Gamma (jj' \lambda i)
\,(\,(-)^{\lambda -\mu }Q_{\lambda \mu i}^{+}\,+\,Q_{\lambda -\mu
i})\,B(jj^{^{\prime }};\lambda -\mu )\,+\,h.c.,\nonumber
\end{equation}
where

\begin{equation}
{\CA} \left( {\lambda i i'} \right)=\frac{X%
^{\lambda i}\,+\,X^{\lambda i^{^{\prime }}}}{\sqrt{Y%
^{\lambda i}Y^{\lambda i^{^{\prime }}}}},\nonumber
\end{equation}

\begin{equation}
\Gamma(jj'\lambda i)=\frac{\pi_{\lambda}}{\pi_{j}} \frac{%
v_{jj'}^{(-)}\,f_{jj'}^{(\lambda )}}{\sqrt{Y^{\lambda i}}},\nonumber
\end{equation}

\begin{equation}
X^{\lambda i}\,=\,\sum_{jj^{^{\prime }}}\,\frac{(f_{jj^{^{\prime
}}}^{(\lambda )}u_{jj^{^{\prime }}}^{(+)})^2\,\varepsilon
_{jj^{^{\prime }}}}{\varepsilon _{jj^{^{\prime }}}^2\,-\,\omega
_{\lambda i}^2},\nonumber
\end{equation}

\begin{equation}
Y^{\lambda i}\,=\,\sum_{jj^{^{\prime }}}\,\frac{(f_{jj^{^{\prime
}}}^{(\lambda )}u_{jj^{^{\prime }}}^{(+)})^2\,\varepsilon
_{jj^{^{\prime }}}\,\omega _{\lambda i}}{\left(\varepsilon
_{jj^{^{\prime }}}^2\,-\,\omega _{\lambda i}^2\right)^2},\nonumber
\end{equation}
with $v^{(-)}_{jj'}=u_{j}u_{j'}-v_{j}v_{j'}$ and
$u^{(+)}_{jj'}=u_{j'}v_{j}+u_{j}v_{j'}$. Here and further we use the
notation $\pi_j=\sqrt{(2j+1)}$.

 The model wave function of an odd spherical nucleus is
taken in the form \cite{Sluys83}:
\begin{equation}\label{eq:oddwv}
\Psi _\nu (JM) = O^{+}_{JM \nu}|\rangle,
\end{equation}
where
\begin{equation}\label{eq:OddCreation}
 O^{+}_{JM \nu}= C_{J\nu}\alpha _{JM}^+ +  \sum_i
D_j^{\lambda i}(J\nu )P_{j\lambda i}^{+}(JM) -
E_{J\nu}\tilde{{\alpha}}_{JM}-\sum_i F_j^{\lambda i}(J\nu
)\tilde{P}_{j\lambda i}(JM),
\end{equation}
with

\begin{equation}
P_{j\lambda i}^{+}(JM) = \left[ \alpha_{jm}^+ Q_{\lambda\mu
i}^{+}\right]_{JM}\nonumber
\end{equation}
and  $\tilde{}$  stands for time conjugate according to the
convention: $\tilde{P}_{j \lambda i} (JM) = (-1)^{J-M}P_{j \lambda
i}(J-M)$.

We apply the equation of motion method to the excitation operator
\eqref{eq:OddCreation}:

\begin{equation}
\langle|\{\delta O_{JM\nu }, H, O_{JM\nu }^{+} \}|\rangle = \eta
_{J\nu}\langle|\{\delta O_{JM}, O_{JM}^{+} \}|\rangle.\nonumber
\end{equation}

Following the linearization procedure \cite{Rowe70}, at the final
state of calculation of the matrix elements, we consider the ground
state to be a vacuum state for both operators $\alpha _{JM}$ and
$Q_{\lambda \mu i}$.

 In all calculations the exact commutation relations
between the quasiparticle and phonon operators are considered:
\begin{equation}
\left[ {\alpha _{jm}^{} ,Q^{+}_{J M \nu}} \right] = \sum\limits_{j'
m' } {\left\langle {{jmj' m' }}
 \mathrel{\left | {\vphantom {{jmj' m' } {J M }}}
 \right. \kern-\nulldelimiterspace}
 {{J M }} \right\rangle \psi _{jj' }^{J \nu} \alpha _{j' m' }^ +
 }.\nonumber
\end{equation}

 The normalization condition of the wave function reads
\begin{align}
& \langle|\{ O_{JM \nu}, O_{JM \nu}^{+} \}|\rangle = C_{J\nu}^2 +
E_{J\nu}^2+\sum_{j \lambda i}[D_j^{\lambda i}(J\nu )]^2 +\sum_{j
\lambda i}[F_j^{\lambda i}(J\nu )]^2 +\nonumber\\
&+ \sum_{j \lambda i j' \lambda' i'} [D_j^{\lambda i}(J\nu
)D_{j'}^{\lambda' i'}(J\nu )+F_j^{\lambda i}(J\nu )F_{j'}^{\lambda'
i'}(J\nu )]{\CL}_{J}(j\lambda i|j'\lambda' i')=1.\nonumber
\end{align}

The equation of motion leads to the following system of linear
equations for each state with quantum numbers $JM$:
\newpage
\begin{equation}
\left( {\begin{array}{*{20}c}
   {\varepsilon _J } & {V\left( {Jj'\lambda ' i' } \right)} & {G\left( J \right)} & { - W\left( {Jj' \lambda ' i' } \right)}  \\
   {V\left( {Jj \lambda i } \right)} & {K_J(j\lambda i | j' \lambda i')} & { - W\left( {Jj \lambda i } \right)} & {S_J\left( {j \lambda i |j' \lambda' i '} \right)}  \\
   {G\left( J \right)} & { - W\left( {Jj' \lambda ' i' } \right)} & { - \varepsilon _J } & { - V\left( {J j' \lambda ' i' } \right)}  \\
   { - W\left( {Jj \lambda  i } \right)} & {S_J\left( {j'\lambda ' i' | j \lambda i } \right)} & { - V\left( {Jj \lambda i } \right)} & {-K_J(j\lambda i | j' \lambda i')}  \\

 \end{array} } \right)\left( \begin{gathered}
  C_{J \nu}  \hfill \\
  D_{j'}^{\lambda ' i'}(J\nu)  \hfill \\
   - E_{J\nu}  \hfill \\
   - F_{j'}^{\lambda ' i'}(J\nu)  \hfill \\
\end{gathered}  \right) = \nonumber
\end{equation}

\begin{equation}\label{eq:EnergyMatrix}
 = \eta _{J\nu} \left( \begin{gathered}
  C_{J\nu}  \hfill \\
  D_{j}^{\lambda i}(J\nu)  + \sum\limits_{j' \lambda' i '} {D_{j'}^{\lambda ' i'}(J\nu) {\CL}_{J}(j\lambda i|j'\lambda' i') }  \hfill \\
   - E_{J\nu}  \hfill \\
   - F_{j}^{\lambda i}(J\nu)  - \sum\limits_{j'\lambda ' i '} {F_{j'}^{\lambda ' i'}(J\nu) {\CL}_{J}(j\lambda i|j'\lambda' i') }  \hfill \\
\end{gathered}  \right)
\end{equation}

 The average value of $H$ over the wave functions \eqref{eq:oddwv} is

\begin{align}
&<\Psi _\nu ^* (JM) H \Psi _\nu (JM)>={}\nonumber   \\
&=[C_{J\nu}^{2}-E_{J\nu}^{2}]\varepsilon_J+ \sum_{j \lambda i j'
\lambda ' i'}(D_j^{\lambda i}(J\nu )D_{j'}^{\lambda ' i '}(J\nu
)+F_j^{\lambda i}(J\nu )F_{j'}^{\lambda ' i '}(J\nu ))I_J(j\lambda
i|j'\lambda' i')\nonumber   \\ &+2\sum_{j \lambda
i}(C_{J\nu}D_j^{\lambda i}(J\nu )-E_{J\nu}F_j^{\lambda i}(J\nu
))V({Jj\lambda i })+2\sum_{j \lambda i}(C_{J\nu}F_j^{\lambda
i}(J\nu)+E_{J\nu}D_j^{\lambda i}(J\nu ))W({Jj\lambda i }) \nonumber
\\&- 2C_{J\nu}E_{J\nu}G_J-2\sum_{j \lambda i j' \lambda '
i'}D_{j}^{\lambda i}F_{j'}^{\lambda ' i '}S_{J}(j \lambda i | j '
\lambda ' i ').\nonumber
\end{align}

We give the explicit expressions for the quantities entering the
above formulas with short comments

\begin{equation}
{\CL}_{J}(j\lambda i|j'\lambda' i') = \pi _\lambda  \pi _{\lambda'}
\sum\limits_{j_1 } {\psi _{j_1 j}^{\lambda'i' }\psi _{j_1
j'}^{\lambda i} \left\{ {\begin{array}{*{20}c}
   {j' } & {j_1 } & \lambda   \\
   {j } & J & \lambda'  \\
 \end{array} } \right\}},\nonumber
\end{equation}

\begin{eqnarray}
G_J = \left\langle {\left| {\left\{ {\tilde \alpha _{J M }^ +
,\left[ {H,\alpha _{J M }^ +  } \right]} \right\}} \right|}
\right\rangle =  \sqrt 2 \sum\limits_{\lambda ij} {\frac{{\pi
_\lambda  }} {{\pi _J }}\Gamma  \left( {J j\lambda i} \right)\varphi
_{J j}^{\lambda i } },\nonumber
\end{eqnarray}

\begin{align}
&V  \left( {Jj\lambda i } \right) = \left\langle {\left| {\left\{
{\left[ {\alpha _{J M }  ,H} \right],P_{j\lambda i }^{ +  }
\left( {J M } \right)} \right\}} \right|} \right\rangle = \nonumber \\
&=  - \frac{1}
{{\sqrt 2 }}\Gamma \left( {Jj\lambda i} \right) \nonumber \\
   &- \frac{1}
{{\sqrt 2 }}\sum\limits_{j' \lambda ' i'} {\left( { {\CT}_J\left( {j
\lambda i;j' \lambda ' i'} \right) + {\CL}_{J}\left( {j \lambda i
|j' \lambda ' i'} \right)} \right)\Gamma \left( {J j' \lambda ' i'}
\right)}.\nonumber
\end{align}

As a result of the application of the equation of motion method, the
matrix elements $V\left( Jj \lambda i \right)$ between quasiparticle
and quasiparticle$\bigotimes$phonon states differ by the ones
obtained earlier \cite{Khuong81} by an additive containing
${\CT}_J\left(j \lambda i | j' \lambda ' i' \right)$
\begin{equation}
{\CT}_{J}(j\lambda i|j'\lambda' i') = \pi _\lambda  \pi _{\lambda'}
\sum\limits_{j_1 } \psi _{j_1 j'}^{\lambda i}{\varphi _{j_1
j}^{\lambda'i' }  \left\{ {\begin{array}{*{20}c}
   {j' } & {j_1 } & \lambda   \\
   {j } & J & \lambda'  \\
 \end{array} } \right\}}.\nonumber
\end{equation}

\begin{align}
&W  \left( {Jj\lambda i} \right) = \left\langle {\left| {\left\{
{\left[ {\alpha _{J M }^{ + } ,H} \right],\tilde P_{j\lambda i} ^{ +
} \left( {J M} \right)} \right\}} \right|}
\right\rangle  = \nonumber \\
&\begin{gathered}
   =  - \frac{1}
{4}\frac{{\pi _\lambda }}
{{\pi _{J } }}\sum\limits_{i'\tau _0 } {{\CA}_{\tau _0 } \left( {\lambda i i'} \right)\varphi _{J j }^{\lambda i'} }  \hfill \nonumber \\
   - \frac{1}
{4}\sum\limits_{\lambda ' j' i' i''\tau _0  } {{\CA}_{\tau _0 }
\left( {\lambda ' i'i''} \right)\frac{\pi _{\lambda '}}{\pi _ J}
\left[ {\varphi _{J j' }^{\lambda' i'} {\CL}_{J}\left( { j \lambda i
|j' \lambda' i''  } \right) - \psi _{J j' }^{\lambda' i''}
{\CT}_{J}\left( {j \lambda i |j' \lambda ' i' } \right)} \right]}.
\hfill \\
\end{gathered}\nonumber
\end{align}

The matrix elements $W\left( Jj \lambda i\right)$ appear after the
inclusion of the backward-going terms in the operator
\eqref{eq:OddCreation} and they present a central issue of this
work.

\begin{eqnarray}
S_J  \left( {j \lambda i |j' \lambda ' i' } \right) = \left\langle
{\left| {\left\{ {\tilde P_{j \lambda i }^{ +  } \left( {J M}
\right),\left[ {H,P_{j' \lambda ' i' }^{ +  } \left( {J M} \right)}
\right]} \right\}} \right|} \right\rangle= - G_{j'} {\CT}_J\left( {
j \lambda i  |j' \lambda ' i' } \right) \nonumber,
\end{eqnarray}

\begin{eqnarray}
 I_J\left( {j \lambda i |j' \lambda ' i' } \right) =
\left\langle {\left| {\left\{ {P_{j \lambda i } \left( {JM}
\right),\left[ {H,P_{j' \lambda ' i' }^{ + } \left( {JM} \right)}
\right]} \right\}} \right|} \right\rangle \nonumber,
\end{eqnarray}

\begin{align}
&I_J\left( {j \lambda i |j' \lambda ' i' } \right) + I_J\left( {j'
\lambda ' i' |j \lambda i } \right) =\nonumber \\
&= 2\delta _{j j' } \delta _{\lambda \lambda ' } \delta _{i i' }
\left( {\omega _{\lambda i }  + \varepsilon _{j } } \right) +
{\CL}_J\left( {j' \lambda ' i' |j \lambda i } \right)\left(
{\varepsilon _{j' j }  + \omega _{\lambda ' i' }  + \omega _{\lambda
i } } \right) - {\CR}_J\left( {j \lambda i  | j' \lambda ' i'  }
\right)\nonumber,
\end{align}

\begin{equation}
 K_J(j\lambda i | j' \lambda i') = \frac{1}
{2}\left[ \begin{gathered}
  I_J\left( {j \lambda i |j' \lambda ' i '} \right) + I_J\left( {j' \lambda ' i ' |j \lambda i } \right)
\end{gathered}  \right] \nonumber
\end{equation}

\begin{align}
&{\CR}_J\left( {j \lambda i |j'\lambda ' i' } \right)=\frac{1}
{4}\sum\limits_{i_1\tau _0 } {\left[ {{\CA}_{\tau _0 } \left(
{\lambda i_1i } \right){\CL}_J  \left( {j' \lambda ' i' |j \lambda
i_1} \right) + {\CA}_{\tau _0 } \left( {\lambda ' i_1i' }
\right){\CL}_J
\left({j \lambda i  |j'\lambda ' i_1} \right)} \right]}+ \nonumber\\
&+ \frac{1} {4}\sum\limits_{\lambda _1 i_1 i_2 j_1 \tau _0 }
{{\CA}_{\tau _0 } \left( {\lambda _1 ii'} \right)\left[ {{\CL}_J
\left( {j \lambda i |j_1 \lambda _1 i_1} \right){\CL}_J  \left( {j'
\lambda ' i' ;j_1 \lambda _1 i_2} \right) + {\CL}_J  \left( {j'
\lambda ' i' |j_1 \lambda _1 i_1} \right){\CL}_J \left( {j \lambda i
|j_1 \lambda _1 i_2} \right)} \right]}\nonumber.
\end{align}

The quantities ${\CL}_{J}(j\lambda i|j'\lambda' i')$ ,
${\CT}_{J}(j\lambda i|j'\lambda' i')$ and ${\CR}_J\left( {j \lambda
i |j'\lambda ' i' } \right)$ vanish if the Pauli principle is not
respected.

\section{Approximations}

\subsection{General}

As has been shown in \cite{Khuong81} $\CL _J$ are alternating
quantities and their diagonal values are much greater than the
nondiagonal ones. This is natural from the physical point of view
as the Pauli principle is violated most probably in the
configurations formed by identical quasiparticles. The same
applies for the new quantities ${\CT} _J$.

\begin{equation}
{\CL}_{J}(j\lambda i|j'\lambda' i') ={\CL}(J j\lambda
i)\delta_{jj'}\delta_{\lambda\lambda'}\delta_{i i'},\nonumber
\end{equation}

\begin{equation}
{\CT}_{J}(j\lambda i|j'\lambda' i') ={\CT}(J j\lambda
i)\delta_{jj'}\delta_{\lambda\lambda'}\delta_{i i'},\nonumber
\end{equation}
where

\begin{equation}
{\CL}(Jj\lambda i) = \pi _\lambda ^2 \sum\limits_{j' } {\left( \psi
_{ j' j }^{\lambda i}\right) ^2 \left\{ {\begin{array}{*{20}c}
   {j } & {j' } & \lambda   \\
   {j } & J & \lambda  \\
 \end{array} } \right\}},\nonumber
\end{equation}

\begin{equation}
{\CT}(Jj\lambda i) = \pi _\lambda ^2  \sum\limits_{j' } \psi _{j'
j}^{\lambda i}{\varphi _{j' j }^{\lambda i }  \left\{
{\begin{array}{*{20}c}
   {j } & {j' } & \lambda   \\
   {j } & J & \lambda  \\
 \end{array} } \right\}}.\nonumber
\end{equation}

In this approximation

\begin{equation}
V  \left( {Jj\lambda i } \right) = - \frac{1} {{\sqrt 2 }}(1 +
{\CL}(J j\lambda i) + {\CT}(J j\lambda i))\Gamma \left( {Jj\lambda
i} \right).\nonumber
\end{equation}

The vertice $V$ is renormalized by the factor $(1 + {\CL}(J j\lambda
i) + {\CT}(J j\lambda i))$. In configurations with strong Pauli
principle violation the quantities ${\CL}(J j\lambda i)$ go to $-1$
and the quantities ${\CT}(J j\lambda i)$ go to $0$. The role of the
term ${\CT}(J j\lambda i)$ in the renormalization becomes more
important as the phonon collectiveness increases.
\begin{eqnarray}
\begin{gathered}
W  \left( {Jj\lambda i} \right) =  - \frac{1} {4}\frac{{\pi _\lambda
}}{{\pi _{J } }}(1+{\CL}(J j\lambda i)-{\CL}(j J\lambda i))\sum\limits_{i'\tau _0 } {{\CA}_{\tau _0 } \left( {\lambda i i'} \right)\varphi _{J j }^{\lambda i'} }  \hfill \nonumber \\
\end{gathered}.\nonumber
\end{eqnarray}

As with the vertices $V$, the vertices $W$ are renormalized now by
the factor $(1+{\CL}(J j\lambda i)-{\CL}(j J\lambda i))$ and again
in configurations with strong Pauli principle violation the
quantities ${\CL}(J j\lambda i)$ go to $-1$ and the quantities
${\CL}(j J\lambda i)$ go to $0$. The results for $V$ and $W$ show
that configurations with ${\CL}(J j\lambda i)$ close to $-1$ must be
excluded from the configuration space.
\begin{align}
K_J\left( {j \lambda i j' \lambda ' i' } \right)= \delta _{j j' }
\delta _{\lambda \lambda ' } \delta _{i i' } \left(1+
{\CL}(Jj\lambda i) \right)\left( {\omega _{\lambda i }  +
\varepsilon _{j } - {\CR} \left( {Jj\lambda i } \right)} \right
)\nonumber,
\end{align}

\begin{equation}
{\CR}  \left( {Jj\lambda i } \right) = \frac{{\CR}_J(j\lambda i|j
\lambda i)}{1+{\CL}(J j\lambda i)}.\nonumber
\end{equation}

The quantities ${\CR}  \left( {Jj\lambda i } \right)$ play a very
important role as they shift the values of the poles and this
shift depends on the extent of the Pauli principle violation
\cite{Khuong81}.

Neglecting $G_J$ and $S_J\left( {j \lambda i |j' \lambda' i '}
\right)$, we arrive at the system of equations \cite{Sluys83}
$$\left[ {\left(
{\begin{array}{*{20}c}
   {\varepsilon _J } & 0  \\
   0 & { - \varepsilon _J }  \\
\end{array}} \right) + \left( {\begin{array}{*{20}c}
   {M_{11} } & {M_{12} }  \\
   {M_{21} } & {M_{22} }  \\
\end{array}} \right)} \right]\left( {\begin{array}{*{20}c}
   {C_{J\nu} }  \\
   { - E_{J\nu} }  \\
\end{array}} \right) = \eta _{J \nu} \left( {\begin{array}{*{20}c}
   {C_{J\nu} }  \\
   { - E_{J\nu} }  \\
\end{array}} \right),\nonumber
$$
where

\begin{align}\label{eq:M11}
&M_{11}  = \sum\limits_{j\lambda i }\frac{1}{\left( {1 + {\CL}\left(
{Jj\lambda i } \right)} \right)} {\left( \frac{{V^2 \left(
{Jj\lambda i } \right)}}{ {\eta _{J \nu}  - \left( {\omega _{\lambda
i } + \varepsilon _j  - {\CR}\left( {Jj\lambda i } \right)} \right)}
} + \frac{W^2 \left( {Jj\lambda i } \right)}{ {\eta _{J \nu}  +
\omega _{\lambda i }  + \varepsilon _j - {\CR}\left( {Jj\lambda i }
\right)} } \right) },
\end{align}

\begin{equation}
M_{22}  = \sum\limits_{j\lambda i } \frac{1}{\left( {1 + {\CL}\left(
{Jj\lambda i } \right)} \right)}{\left( {\frac{{W^2 \left(
{Jj\lambda i } \right)}}{{ {\eta _{J \nu}  - \left( {\omega
_{\lambda i } + \varepsilon _j  - {\CR}\left( {Jj\lambda i }
\right)} \right)} }} + \frac{{V^2 \left( {Jj\lambda i} \right)}}{{
{\eta _{J \nu} + \omega _{\lambda i }  + \varepsilon _j  -
{\CR}\left( {Jj\lambda i } \right)} }}} \right)},\nonumber
\end{equation}

\begin{align}
&M_{12}  = M_{21}  = \nonumber\\
 &=\sum\limits_{j\lambda
i}{\frac{V(Jj\lambda i)W(Jj\lambda i)}{(1 + {\CL}(Jj\lambda
i))}\left(\frac{1}{\eta _{J \nu}+ \omega _{\lambda i } +
\varepsilon_j - {\CR}( Jj\lambda i )}-\frac{1}{\eta _{J \nu}-\left(
\omega _{\lambda i } + \varepsilon_j - {\CR}\left( Jj\lambda i
\right)\right)}\right)} \nonumber,
\end{align}

leading to the equation
\begin{equation}\label{eq:OddAppr1}
{M_{12} M_{21}  = \left( {\varepsilon _J  + M_{11}  - \eta _{J \nu}
} \right)\left( {M_{22}  - \varepsilon _J  - \eta _{J \nu} }
\right)}.
\end{equation}

\subsection{Limit cases and analisys}

The equation \eqref{eq:OddAppr1} can be approximated by the
following one
\begin{equation}\label{eq:OddAppr2}
 \varepsilon _J  - \eta_{J \nu}  \approx -M_{11}-\frac{M_{12}^{2}}{|2\varepsilon _J - \left( M_{22} -M_{11}
 \right)|}.
\end{equation}

Therefore, neglecting the backward amplitudes, i.e. setting $W\left(
{Jj\lambda i } \right) = 0$, \eqref{eq:OddAppr2} immediately reduces
to the secular equation obtained earlier \cite{Khuong81} :
\begin{equation}\label{eq:OddAppr3}
 \varepsilon _J  - \eta_{J \nu}  = \sum\limits_{j \lambda i } {\frac{{V^2 \left( {J j
\lambda i } \right)}} {{\left( {\left( {\omega _{\lambda i }  +
\varepsilon _j  - {\CR}\left( {J j \lambda i  } \right)} \right) -
\eta _{J \nu} } \right)\left( {1 + {\CL}\left( {J j \lambda i  }
\right)} \right)}}}.
\end{equation}

 The significant difference for the solution $\eta _J$ of the
equation \eqref{eq:OddAppr2} as compared to the equation
\eqref{eq:OddAppr3} comes from the second term in the r.h.s. of the
expression \eqref{eq:M11} which contributes to a shift of the first
solution of equation \eqref{eq:OddAppr2} to higher energies. The
second term of the r.h.s. of equation \eqref{eq:OddAppr2} also
contributes in the same direction but to a much smaller extent. The
shift in energy becomes larger as the interaction between the
quasiparticles and phonons increases. A critical value for the
interaction exists as in \eqref{eq:OddAppr3} but now due to the
second type of terms in $M_{11}$ an increase in the interaction
leads to a shift of the first solution towards the pole as contrary
to the case neglecting backward amplitudes where the solution moves
in the opposite direction.

 The inclusion of the matrix element $G_J$ in the system of
 equations can be accounted as a change of $M_{12}$ to $M_{12}+G_J$ but
 the resulting shift in the first solution turns out to be negligible.

\section{Numerical results}
In order to give a qualitative picture of the effects on the
structure of the low-lying states imposed by the backward-going
amplitudes, numerical calculations for $^{131}$Ba were performed.
This isotope belongs to the transitional region where the anharmonic
effects play a gradually increasing role at low and mainly at
intermediate energies and therefore the results presented in this
section may lack some accuracy because the wave function
\eqref{eq:oddwv} does not contain configurations to account for
these effects. The parameters of the Woods-Saxon potential are as
follows:

\begin{center}
\begin{tabular} {|c|c|c|c|c|c|c| } \hline
 $A$ & $N,Z$ & $r_0$, fm & $V_0$, MeV & $\kappa$, fm$^{-2}$ &$\alpha$, fm$^{-1}$ & $G_{N,Z}$, MeV \\ \hline
 $127$ & $N=74$ & 1.280 & 43.40 & 0.413 & 1.613 & 0.124 \\
 $127$ & $Z=53$ & 1.240 & 59.72 & 0.350 & 1.587 & 0.130 \\
 \hline
\end{tabular}
\vspace{2mm}

 {\scriptsize { Table 1. Parameters of the Woods-Saxon
potential for $^{131}$Ba}}
\end{center}

\vspace{2mm} Our study shows that in realistic calculations one must
include phonons with $\lambda = 2, 3, 4, 5$.
 The strength parameters $\kappa^{(\lambda)}$ are adjusted so that the odd energy spectrum of
the low-lying states is reasonably close to the experimental
values. As a result, the energy of the first quadrupole  state of
$^{130}$Ba has a value that is  much higher than the experimental
one.

\begin{figure}
\begin{center}
\includegraphics[width=\textwidth]{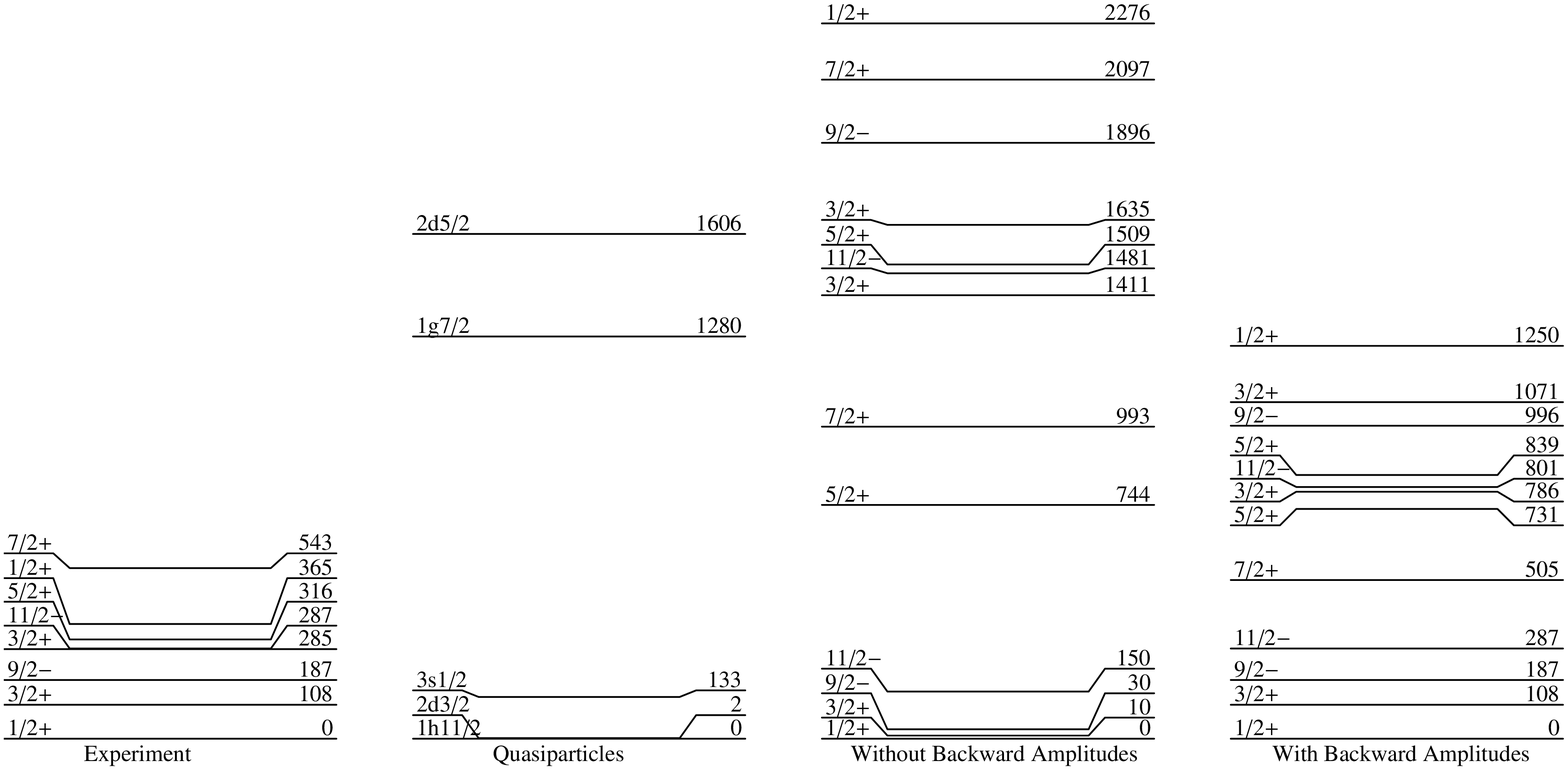}
\end{center}
\scriptsize{Low-lying
energy spectrum of $^{131}$Ba (in KeV). The first column gives the
experimental values \cite{Nucle94}, the second is the unperturbed
1qp spectrum, the third gives the levels resulting from the solution
of \eqref{eq:OddAppr3}, the fourth is the full calculation with
backward amplitudes and Pauli principle corrections. }
\end{figure}

 Solving the systems of equations
\eqref{eq:EnergyMatrix}, one can find the structure of the wave
 functions \eqref{eq:oddwv} and the energies of
the excited states. Working in a diagonal approximation for $\CL _J$
and $\CT _J$ this system reduces to a generalized eigenvalue
problem. In fig.1 a comparison between the experimental values of
the energies and the theoretical calculations is presented.
 We restrict the calculation to the six
neutron states ${1/2}^+,{3/2}^+,{9/2}^-,{11/2}^-,{5/2}^+,{7/2}^+$.
The level ordering presented in the third column on this figure
generally agrees with the one obtained in \cite{Aliko81}. The
results clearly support the conclusion following from
\eqref{eq:OddAppr2} as the first solutions obtained after the
inclusion of the backward-going terms become closer to the first
poles and consequently closer to the second solution thus
significantly reducing the gap between the first $1/2^+$ and the
second $1/2^+$ states as well as between the first $3/2^+$ and the
second $3/2^+$ states. The intruder state ${9/2}^-$ deserves a
special attention. The wave function of this state is practically a
pure quasiparticle$\bigotimes$phonon state with a structure
$[1h_{11/2}\bigotimes2_1^+]_{9/2^-}$. The significant reduction of
the energy of this state is due to the Pauli principle correction,
the inclusion of which is essential for the correct ordering of the
first several levels. Along with the experimental energies our
calculations provide a reasonable description of the spectroscopic
factors for the $(d,p)$-reactions: for the states $1/2+$ and $3/2+$
having experimentally measured values of the spectroscopic factors
of $0.53$ and $1.03$ respectively, our calculations give $0.46$ and
$1.36$ .

\section{Conclusion}

The comparison between theoretical calculations and experimental
data for $^{131}$Ba has shown that in order to describe the
structure of the low-lying states in odd-mass nuclei far from the
magic numbers one needs to take into account the Pauli principle and
the ground state correlations effects simultaneously. Calculations
for other Ba isotopes are in progress now. To improve this approach
a self-consistent description of the mean field with more realistic
effective nucleon-nucleon forces is desirable.

\end{document}